\documentclass{iau}
\usepackage{graphicx}

\title[YSOs in Canis Major] 
{The WISE Census of Young Stellar Objects in Canis Major}

\author[W. J. Fischer, D. L. Padgett, \& K. R. Stapelfeldt]   
{W. J. Fischer$^{1,2}$, D. L. Padgett$^1$, \and K. R. Stapelfeldt$^1$}

\affiliation{$^1$NASA Goddard Space Flight Center\\email: {\tt william.j.fischer@nasa.gov}\\
$^2$NASA Postdoctoral Program Fellow}

\pubyear{2015}
\volume{314}  
\jname{Young Stars \& Planets Near the Sun}
\editors{J. H. Kastner, B. Stelzer, \& S. A. Metchev, eds.}
\begin{document}

\maketitle

\begin{abstract}
While searches for young stellar objects (YSOs) with the Spitzer Space Telescope focused on known molecular clouds, photometry from the Wide-field Infrared Survey Explorer (WISE) can be used to extend the search to the entire sky.  As a precursor to more expansive searches, we present results for a 100 deg$^2$ region centered on the Canis Major clouds.
\end{abstract}

\firstsection 
\section{Introduction}

With its all-sky survey at 3.4, 4.6, 12, and 22 $\mu$m, the Wide-field Infrared Survey Explorer (WISE; \cite[Wright \etal\ 2010]{wri10}) can be used to identify young stellar objects (YSOs) with criteria similar to those established for the Spitzer Space Telescope but over the entire sky.  Newly identified YSOs may refine the initial stellar mass function, allow a better characterization of star and planet formation in regions with low initial gas densities, and identify nearby targets for high-resolution follow-up imaging.

As a pilot study for our more expansive search, we present results for Canis Major.  Star formation in the vicinity of Canis Major is centered amid the CMa OB1 association, 1--2$^\circ$ below the Galactic plane near a longitude of $224^\circ$ and at a distance of $\sim$ 1000 pc (\cite[Gregorio-Hetem 2008]{gre08}).  It is home to $5\times10^4$ $M_\odot$ of material distributed across 22 clouds as traced by $^{13}$CO gas (\cite[Kim \etal\ 2004]{kim04}).  Although parts of it have recently been mapped by the outer Galactic plane surveys of Spitzer and the Herschel Space Observatory, there is no comprehensive study of star formation across the region.  We searched the AllWISE catalog for Class I YSOs, in which a dusty protostellar envelope dominates the infrared emission, and more evolved Class II YSOs, in which a dusty circumstellar disk dominates.

\section{YSO Selection Techniques}

We targeted the $10^\circ \times 10^\circ$ square centered at $106.67^\circ$ right ascension and $-11.29^\circ$ declination.  For inclusion in our initial catalog, we required detections in Bands 1 and 2, and we ignored sources flagged as artifacts or extended emission.  To identify candidates, we adopted the WISE color-color criteria of \cite[Koenig \& Lesiawitz (2014)]{koe14}.  These are based on the colors of known YSOs in Taurus, extragalactic sources, and galactic contaminants.

Most of the sources selected by these color criteria are faint and uniformly distributed across the region, unlike YSOs, which should be clustered near the sites of their formation.  To deselect these likely extragalactic sources, we required a WISE band 1 magnitude $W1<12$ or a WISE band 4 magnitude $W4<5$. To remove bright galactic contaminants such as red giant stars, we required $W1>6$.  Of the sources with the requisite colors, 155 Class I and 375 Class II candidates satisfy the magnitude requirements in the full 100 deg$^2$ search region.

\section{A New Young Cluster of YSOs}

\begin{figure}
\includegraphics[width=\hsize]{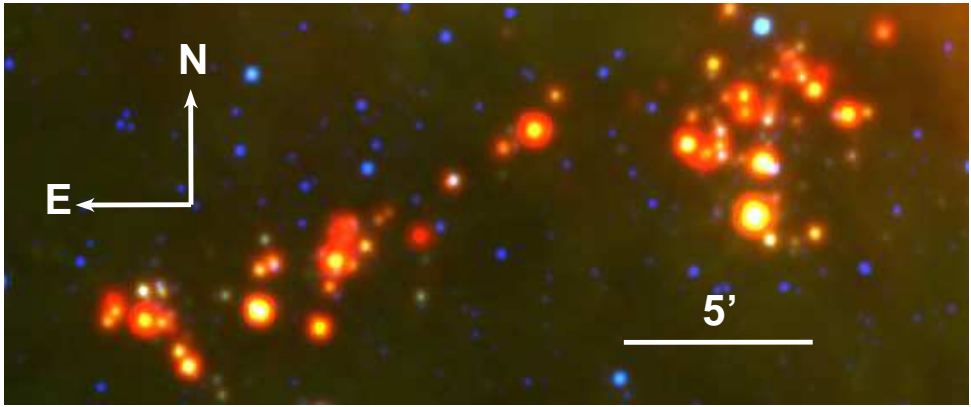}
\caption{A three-color WISE image of a YSO cluster in Canis Major, centered at $7^h 10^m$ RA and $-10^\circ30'$ Dec.  Blue is 3.6 $\mu$m, green is 12 $\mu$m, and red is 22 $\mu$m.  The $5'$ scale bar corresponds to 1.45 pc at a distance of 1000 pc. Generally, the bright red sources are the Class I candidates.  {\em See the electronic edition for full color.}\label{fig:wise}}
\end{figure}

We discovered a cluster of YSOs (Fig.\,\ref{fig:wise}) located 2$^\circ$ to the northeast of the well studied object Z CMa.  Assuming a distance of 1000 pc, it covers an area of 9.3 pc$^2$ according to the method of \cite[Gutermuth \etal\ (2009)]{gut09}.  The cluster contains 31 Class I candidates based on the above criteria, 20\% of the total for the 100 deg$^2$ search region.  It also contains 12 Class II candidates, although this is a lower limit due to the insensitivity of WISE bands 3 and 4 to sources with Class II colors and magnitudes at 1000 pc.   This cluster has more Class I sources and a larger physical area than about 80\% of the 36 clusters within 1~kpc of the sun studied by \cite[Gutermuth \etal\ (2009)]{gut09}, and the density of Class I sources is 3.3~pc$^{-2}$, larger than that of about two thirds of the Gutermuth sample.

The reddest source, just east of the center of Figure~\ref{fig:wise}, has WISE magnitudes consistent with a 4.5 $L_\odot$ Class 0 protostar at 1000 pc.  Its detection suggests that the WISE 22~$\mu$m channel is useful for identifying isolated, deeply embedded young protostars.

\section{Conclusions}

With WISE photometric criteria, we detected over 500 candidate YSOs in a $10^\circ \times 10^\circ$ square centered on the Canis Major star forming region.  The most populous cluster of candidates has 31 Class I candidates and an area of 9.3 pc$^2$, typical of the larger, more populous clusters in the nearest 1 kpc.  Identification of analogs to the nearby star forming regions at larger distances provides the opportunity for efficient environmental studies with current and future long-wavelength observing facilities.  Subsequent application of our YSO finding criteria to the entire sky may reveal isolated young stellar objects near the sun for high-resolution studies.

\end{document}